\documentclass[sigconf]{acmart}

\AtBeginDocument{%
  }

\acmYear{2024}\copyrightyear{2024}
\acmConference[MOBIHOC '24]{The Twenty-fifth International Symposium on Theory, Algorithmic Foundations, and Protocol Design for Mobile Networks and Mobile Computing}{October 14--17, 2024}{Athens, Greece}
\acmBooktitle{The Twenty-fifth International Symposium on Theory, Algorithmic Foundations, and Protocol Design for Mobile Networks and Mobile Computing (MOBIHOC '24), October 14--17, 2024, Athens, Greece}
\acmDOI{10.1145/3641512.3690630}
\acmISBN{979-8-4007-0521-2/24/10}

\begin{document}

\title{Demo: FedCampus: A Real-world Privacy-preserving Mobile Application for Smart Campus via Federated Learning \& Analytics}




\author{Jiaxiang Geng \quad Beilong Tang \quad Boyan Zhang \quad Jiaqi Shao \quad Bing Luo}
\affiliation{
    \institution{Duke Kunshan University, Kunshan, Jiangsu, China}
    \country{}
}
\email{{jg645,bt132,bz106,js1139,bl291}@duke.edu}

\renewcommand{\shortauthors}{Jiaxiang Geng et al.}

\begin{abstract}
In this demo, we introduce FedCampus, a privacy-preserving mobile application for smart \underline{campus} with \underline{fed}erated learning (FL) and federated analytics (FA). FedCampus enables cross-platform on-device FL/FA for both iOS and Android, supporting continuously models and algorithms deployment (MLOps). 
Our app integrates  privacy-preserving processed data via differential privacy (DP) from smartwatches, where the processed parameters are used for FL/FA
through the FedCampus backend platform.
We distributed 100 smartwatches to volunteers at Duke Kunshan University and have successfully completed a series of smart campus tasks featuring capabilities such as sleep tracking, physical activity monitoring, personalized recommendations, and heavy hitters. Our project is opensourced at https://github.com/FedCampus/FedCampus\_Flutter.
See the FedCampus video at https://youtu.be/k5iu46IjA38.
\end{abstract}



\thanks{This work is supported by Suzhou Frontier Science and Technology Program (Project SYG202310) and Kunshan Municipal Government research funding (Project 24KKSGR013). We acknowledge Sichang He, Qingning Zeng, Luyao Wang, and Renyuan Zhang for their contributions to this work.  Corresponding Author: Bing Luo.}


\maketitle

\section{Introduction}
Federated Learning (FL) holds great promise for collaboratively training shared machine learning models on end devices while preserving data privacy\cite{cost}. 
However, apart from Google's keyboard prediction and Apple's speaker recognition in Siri, there are few real-world applications of FL, especially for cross-platform on-device scenarios. Although there are FL frameworks such as Flower\cite{flower} and FedML\cite{fedml}, they have not been deployed as real-world applications.

FedCampus address the following challenges: 1) We develop and launch a real-world privacy-preserving mobile application at a university scale.
2) To collaboratively train the same models across users' diverse smartphones, we address cross-platform on-device training and model aggregation. 3) To customize FL algorithms and update models in production, we develop to get maximal control over the FL process on users' devices from our end. 4) To enhance privacy protection during real-world applications, we support encryption features such as differential privacy (DP) on mobile devices, including encryption in both FL and Federated Analytics (FA)\footnote{Unlike FL, FA clients send encrypted parameters to the server for analysis instead of training models on edge devices, making it suitable for scenarios with limited computational capability on edge devices.}.  


\begin{figure} \centering
  {\includegraphics[width=1\linewidth]{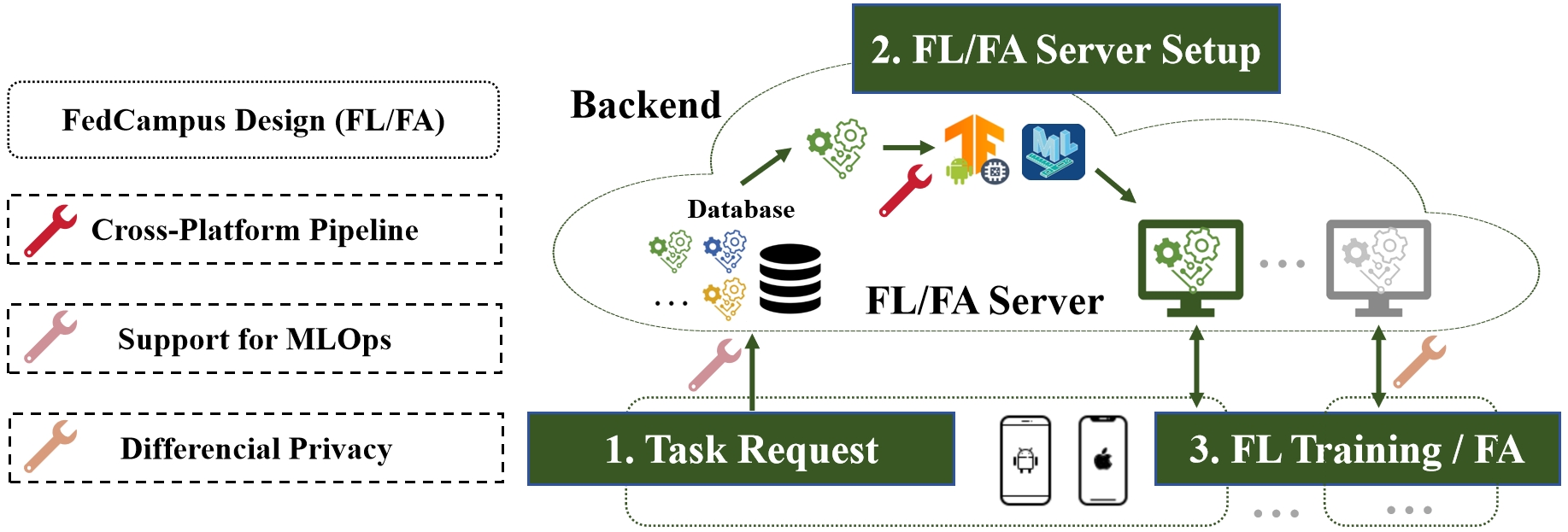}}
 \vspace{-8mm}
 \caption{FedCampus Workflow.} \label{fig:FedCampus Workflow}
 \vspace{-5mm}
\end{figure}

\vspace{-2mm}

\section{FedCampus Design}
FedCampus enables FL/FA across both Android and iOS client devices, managed by a single backend server, as shown in Fig.\ref{fig:FedCampus Workflow}. For FL, each client trains a local model using private data, while the backend server aggregates these local models across platforms to update the global model.

\vspace{-2mm}

\subsection{Cross-Platform On-Device FL Pipeline}
To enable cross-platform FL \cite{he2024fedkit}, particularly cross-platform aggregation, we propose a pipeline that includes model conversion and unified training APIs:

\textbf{Model Conversion: } We convert models into formats compatible with Android (TensorFlow Lite or TFLite) and iOS (Core ML). For Android, we standardize a model format and develop a compliant TensorFlow converter. For iOS, we define a fixed model structure and utilize the official CoreMLTools converter.

\textbf{Unified Training APIs: } FedCampus provides trainers with unified APIs for retrieving and setting model parameters, model fitting, and evaluation, leveraging GPU and NPU acceleration. On Android, we use the TFLite interpreter to invoke our standardized methods defined during model conversion. On iOS, we employ an undocumented method that modifies the underlying ProtoBuf representations of Core ML models.

\textbf{Cross-Platform Aggregation: } Aggregation requires uniform parameter representations, which presents challenges in retrieving and setting parameters. On Android, we resolve the issue of the TFLite interpreter only accepting inputs/outputs as name-to-tensor maps by using index-based methods to access parameters. On iOS, we use ProtoBuf manipulation to manage Core ML’s restrictions on updatable layers and post-training parameter access.

\begin{figure}[] \centering
  \centering
 \includegraphics[width=0.9\linewidth]{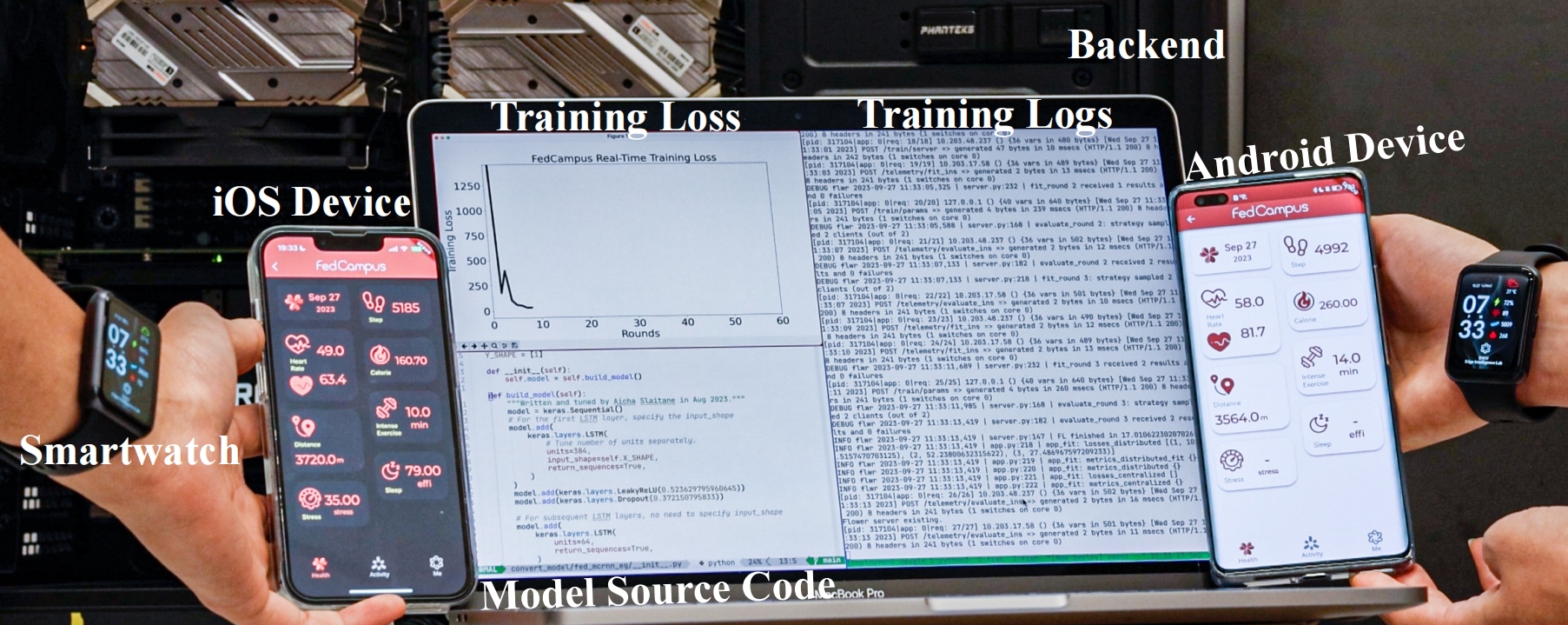}
 \vspace{-4mm}
    \caption{FedCampus Setup}
    \label{fig:architecture}
\vspace{-4mm}
\end{figure}

\vspace{-2mm}

\subsection{Support for MLOps}
In production, FL development faces challenges due to the lack of direct control over end devices. FedCampus addresses this by empowering researchers to continuously deploy models and algorithms (MLOps).

\textbf{Continuous Cross-Platform Task Delivery: }
FedCampus enables continuous task delivery without requiring app updates by decoupling models from clients through a Model Request system. This allows new models to be deployed by uploading them to the Backend. 

\textbf{Customizable Continuous FL Training: } FedCampus manages continuous FL training by supporting multiple parallel training sessions through FL Server Setup. Each FL Server runs as an independent Python subprocess within the Django Backend, occupying its own port for client connections during FL Training. 

\vspace{-2mm}

\subsection{Differential Privacy on FL \& FA}
Although FL and FA inherently offer privacy-preserving features, there are still privacy risks in production, such as exposure to membership inference attacks. To enhance privacy protection, FedCampus provides interfaces for privacy-preserving algorithms, enabling the implementation of DP and other encryption methods.

\textbf{Interface for DP on FL \& FA: } FedCampus supports both FL and FA, applying DP encryption as an example in both contexts: For FL, FedCampus adds noise to models after local training on client devices and supports the aggregation of these noisy models on the server. For FA, FedCampus implements de-identification of local data on devices and applies noise perturbation to the data, which effectively protects client data privacy during analysis and communication.

\vspace{-2mm}

\section{Demonstration}
FedCampus has been deployed and validated in real-world environments. We recruited a large number of volunteers and implemented the system across various personal devices, successfully completing multiple smart campus tasks in Duke Kunshan University.

\vspace{-2mm}

\subsection{Demonstration Setup}
Here, we outline our setup implemented for FedCampus in Duke Kunshan University, as shown in Fig.\ref{fig:architecture}. The platform has been operational for 11 months, from 2023/10 to the present.

\textbf{Participants: }  A total of 100 campus volunteers participated in the testing, each equipped with a wearable device configured to automatically record daily health data such as average heart rate, steps, calorie and sleeping time. The volunteers came from diverse backgrounds and exhibited varying levels of physical activity. All the privacy-sensitive health data were anonymized by removing identifiable user information and further protected through our supported DP techniques.

\textbf{Hardware: } Client hardware included iOS devices, Android devices, and Huawei smartwatches. Server hardware comprised a DELL PowerEdge T640. The system utilized LTE networks, 5G networks, and campus Wi-Fi networks for communication.

\textbf{Software: } The software components included FedCampus Android and iOS clients' app and a server backend. We also integrated the Huawei Health Kit. Personal health data from smartwatch is sent to the mobile phone, and then the mobile phone connects with the server to enable FL/FA.

\begin{figure}[] \centering
  \centering
 \includegraphics[width=\linewidth]{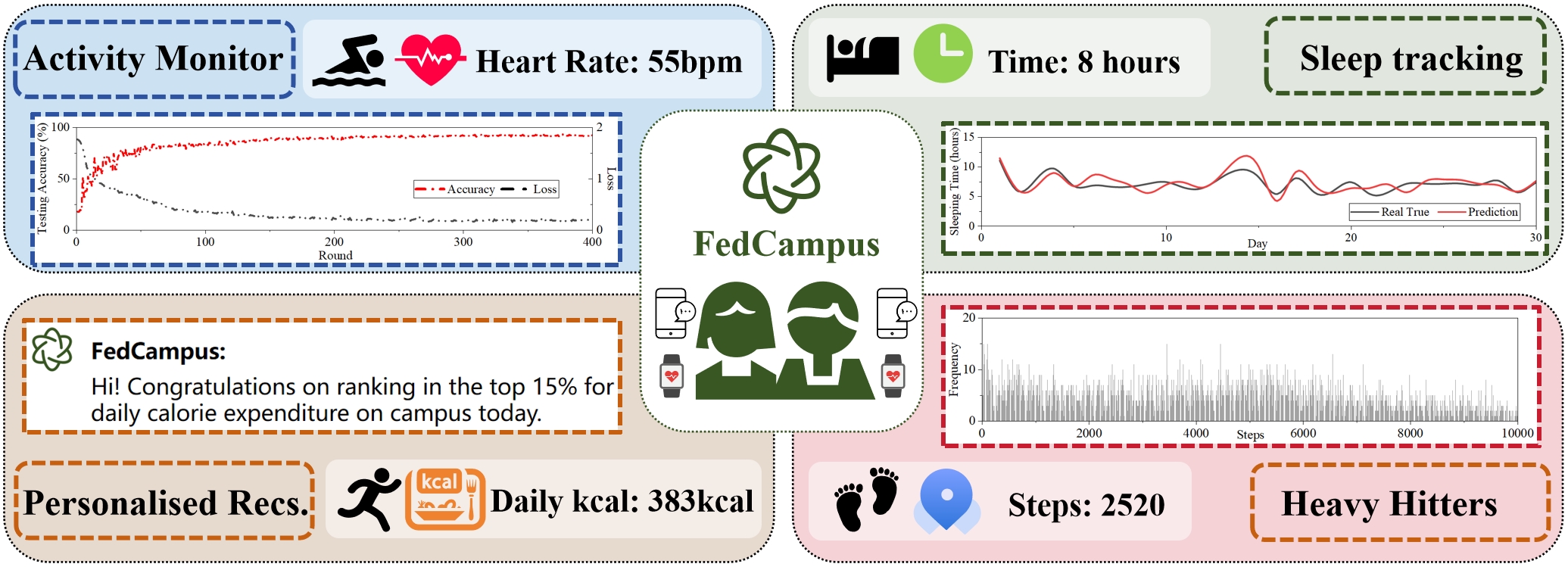}
 \vspace{-8mm}
    \caption{Smart Campus Tasks}
    \label{fig:task}
\vspace{-4mm}
\end{figure}

\vspace{-1mm}

\subsection{Smart Campus Tasks}
FedCampus has successfully implemented a variety of smart campus tasks, as shown in Fig.\ref{fig:task}.

\textbf{Sleep Tracking (FL): } This task monitors and analyzes sleep patterns using labels such as sleeping duration and screen time to provide insights into sleep quality. We have trained a sleep-efficiency prediction model \cite{sleep}.

\textbf{Physical Activity Monitoring (FL): } This task tracks and evaluates physical activities using labels like steps, walking distance, and average heart rate to assess overall fitness levels. We have trained a CNN model for activity prediction \cite{fedhome}.

\textbf{Personalized Recommendations (FA): } This task delivers tailored suggestions to users based on aggregated health data, utilizing labels like activity levels and daily calorie.

\textbf{Heavy Hitters (FA): } Our target is to identify the most prevalent step counts, known as "heavy hitters," across various participant clusters to evaluate the quality of students' physical activity \cite{shao2023privacy}.


\bibliographystyle{ACM-Reference-Format}
\bibliography{citations}

\end{document}